\documentclass[prd,showpacs,preprintnumbers,onecolumn,superscriptaddress,amsmath,nofootinbib,amssymb]{revtex4}
\usepackage{graphicx,color,dcolumn,booktabs,bm}
\usepackage{longtable,lscape}
\usepackage{txfonts}
\usepackage{overpic}
\usepackage{amssymb}
\usepackage{epstopdf}
\usepackage{indentfirst}
\usepackage{feynmf}   
\usepackage{slashed}  
\usepackage{cases}
\usepackage{color}
\usepackage{float}
\usepackage{multirow}
\usepackage{graphicx,color,dcolumn,booktabs,bm}
\usepackage{epsfig,dsfont,amssymb,amsmath,amsfonts,amsbsy,mathrsfs}
\usepackage{mathrsfs}

\graphicspath{{Figures/}} %

\usepackage{hyperref}
\hypersetup{colorlinks,citecolor=blue,anchorcolor=red,menucolor=red, linkcolor=red,filecolor=red,runcolor=red,urlcolor=blue,frenchlinks=red}

\makeatletter
\@addtoreset{equation}{section}
\makeatother

\allowdisplaybreaks

\begin{document}

\title{Study of the radiative decay $J/\psi\rightarrow\eta_{c}+\gamma$ in light cone sum rules}

\author{Song-Pei Guo}
\email{guosongpei6235@163.com}
\affiliation{College of Physics and Electronic Engineering, Northwest Normal University, Lanzhou 730070, China
}
\affiliation{Research Center for Hadron and CSR Physics,
Lanzhou University and Institute of Modern Physics of CAS, Lanzhou 730000, China}

\author{Yan-Jun Sun}
\email{sunyanjun@nwnu.edu.cn}
\affiliation{College of Physics and Electronic Engineering, Northwest Normal University, Lanzhou 730070, China
}
\affiliation{Research Center for Hadron and CSR Physics, Lanzhou University and Institute of Modern Physics of CAS, Lanzhou 730000, China}
\author{Wei Hong}
\email{hongwei17809213575@163.com}
\affiliation{College of Physics and Electronic Engineering, Northwest Normal University, Lanzhou 730070, China
}
\affiliation{Research Center for Hadron and CSR Physics, Lanzhou University and Institute of Modern Physics of CAS, Lanzhou 730000, China}
\author{Qi Huang}
\email{qihuang1193572279@163.com}
\affiliation{College of Physics and Electronic Engineering, Northwest Normal University, Lanzhou 730070, China
}
\affiliation{Research Center for Hadron and CSR Physics, Lanzhou University and Institute of Modern Physics of CAS, Lanzhou 730000, China}
\author{Guo-Hua Zhao}
\email{guohuazhao0916@163.com}
\affiliation{College of Physics and Electronic Engineering, Northwest Normal University, Lanzhou 730070, China
}
\affiliation{Research Center for Hadron and CSR Physics, Lanzhou University and Institute of Modern Physics of CAS, Lanzhou 730000, China}
\begin{abstract}
At present, there are many experimental and theoretical results for $J/\psi\rightarrow\eta_{c}+\gamma$ process, while the results are inconsistent based on different methods and considerations. In this paper, the light cone sum rules method is used in studying the radiative decay $J/\psi\rightarrow\eta_{c}+\gamma$. We give the transition form factor of this process based on the leading twist distribution amplitude of $\eta_{c}$ meson, with this form factor further obtain the decay width. Our result is consistent with those of other sum rules. A comparison of our result with others' about the decay width is also presented.

\end{abstract}


\maketitle

\section{introduction}\label{sec1}

   Radiative transition plays an important role in the establishment of new quarkonium states for a long time, and many theoretical and experimental studies have been carried out in this area.
   Since the photon in radiative transition serves as a clean probe into inner structure of the  quark-antiquark bound states, a thorough understanding of radiative transition
  process will enrich our knowledge of underlying dynamics.

 Radiative transition in heavy quarkonium is similar to electromagnetic transitions of ``hydrogen atom", therefore development of many theoretical machineries in atomic physics can be directly borrowed. In this paper, $J/\psi\rightarrow\eta_{c}+\gamma$ decay is a magnetic dipole radiative transition between charmoniums through the emission of a photon. The magnetic dipole radiative transition flips the spin of one of heavy quarks without changing the orbital angular momentum.

   Comparing the experimental and theoretical rates of radiative transition offers a guidance to enhance our understanding of the internal structure of charmonium. Over years, some radiative transition experiments involving $J/\psi\rightarrow\eta_{c}+\gamma$ decay have been measured~\cite{Gaiser:1985ix,Mitchell:2008aa,Anashin:2014wva}, and comprehensive theoretical studies have been carried out. The method of dispersion sum rules~\cite{Khodjamirian:1979fa} for $J/\psi\rightarrow\eta_{c}+\gamma$ decay was used on the basis of local duality hypothesis in early papers, however, the local duality hypothesis for the lowest state is not merely a consequence of general principles, because if it is true, one should be proved on the basis of specific dynamics of the process considered. Beilin~\cite{Beilin:1984pf} analyzed the radiative decay with the QCD sum rules approach including both nonperturbative $O(\langle GG\rangle)$ and perturbative $O(\alpha_{s})$ corrections, and took appropriate ratio of 3-point function to 2-point function moments. Later, relativistic quark model~\cite{Ebert:2002pp}, constituent quark model~\cite{Deng:2016stx}, non-relativistic potential model~\cite{Barnes:2005pb}, and coulomb gauge approach~\cite{Guo:2014zva} have been employed. However, it is not clear what is the relationship between phenomenological model and the first principle of QCD. The nonrelativistic effective field theories of QCD~\cite{Brambilla:2005zw,Pineda:2013lta} based on the effective field theory (EFT) method has become available, but Ref.~\cite{Brambilla:2005zw} clearly indicates that the transition is sensitive to relativistic effect. Furthermore, lattice QCD~\cite{Dudek:2006ej,Dudek:2009kk,                                     Gui:2019dtm,Donald:2012ga,Chen:2011kpa,Becirevic:2012dc} have also emerged.

  It is clear that the sum rules method is based on quantum chromodynamics (QCD), while the nonrelativistic effective field theories of QCD lacks relativistic effect. Therefore, the light cone sum rules method on the basis of QCD and relativistic effect is appropriate in this work, because the wave function in this approach provides a reasonable description of the solution of relativistic bound state equation. The method of light-cone sum rules is a fruitful hybrid of the SVZ sum rules~\cite{Shifman:1978bx,Colangelo:2000dp} technique and the theory of hard exclusive processes. In the present work we calculate the form factor and the decay width for the radiative transition $J/\psi\rightarrow\eta_{c}+\gamma$ within light-cone sum rules approach. In our previous work, this method has been successfully applied to $e^{+}+e^{-}\rightarrow J/\psi +\eta _{c}$ process~\cite{Sun:2009zk} and $e^{+}+e^{-}\rightarrow \psi(2S)+\eta _{c}$ process~\cite{Tao:2019rwy}. After careful analysis of the hadronic spectra, we find that the light-cone sum rules method can also be used for this radiative transition process. This paper is structured as follows. In Section II, we derive the transition form factor and the corresponding decay width at $Q^{2}=0$  in the framework of the light-cone sum rules. Our numerical analysis and discussion about form factor and decay width of this process are presented in Section III. Finally, a summary is given in Section IV.

\section{The form factor and the decay width }\label{sec2}
A.Transition form factor at $Q^{2}=0$

First we study the electromagnetic form factor involved in the radiative decay $J/\psi(P)\rightarrow\eta_{c}(P^{\prime})+\gamma(q)$ at $Q^{2}=0$ (for real photon). To achieve this goal, the two-point correlation function between vacuum and on-shell state can be constructed as,
\begin{eqnarray}
\label{eq9}
\Pi_{\mu \nu }\left ( P^{\prime},q \right )=i\int d^{^{4}}xe^{iqx}\langle\eta _{c}\left ( P^{\prime} \right )\mid T\left \{ J_{\mu }^{c}\left ( x \right )J_{\nu }^{c}\left ( 0 \right ) \right \}\mid 0\rangle,
\end{eqnarray}
in Eq.(\ref{eq9}), $q$ is the photon momentum, $P^{\prime}$ stands for the four-momentum of $\eta _{c}$ meson, and $J_{\mu }^{c}$ is the colorless c-quark vector current,
\begin{eqnarray}
J_{\mu }^{c}(x)=\bar{C}\left ( x \right )\gamma _{\mu }C\left ( x \right ).
\end{eqnarray}

 The transition form factor can be obtained by calculating the correlation function from hadronic and operator product expansion (OPE) sides and then matching the results on the two sides. On the hadronic side which is in time-like region, we insert complete sets of intermediate states with the same quantum numbers as the $J/\psi $ ground state between the two electromagnetic currents, and separate the contribution of the $J/\psi $ ground state from the excited and continuum states. As a result, the hadronic part of the correlation (\ref{eq9}) can be written as,
\begin{eqnarray}
\label{eq1}
\Pi _{\mu \nu }\left ( P^{\prime},q \right )&=&\frac{1}{m_{J/\psi }^{2}-\left ( P^{\prime}+q \right )^{2}}\langle\eta _{c}\left ( P^{\prime} \right ) \mid J_{\mu }^{c}\left ( 0 \right )\mid  J/\psi \left ( P^{\prime} +q\right )\rangle \langle J/\psi \left (P^{\prime} +q\right )\mid J_{\nu }^{c}\left ( 0 \right )\mid 0\rangle\nonumber\\
&&+\frac{1}{\pi }\int_{s_{0}}^{\infty }ds\frac{\rm{Im}\Pi _{\mu \nu }}{s-\left ( P^{\prime}+q \right )^{2}},
\end{eqnarray}
where $m_{J/\psi }$ is the $J/\psi $ mass, $s_{0}$ is the threshold parameter separating the ground state from the excited states and continuum. The first term represents the contribution from $J/\psi $, and the second denotes the dispersion integral that includes the contributions of the excited states and the continuum. The coupling between $J/\psi$ and the electromagnetic current  $J_{\mu }^{c}$ is defined as
\begin{eqnarray}
\label{eq11}
\langle0\mid J_{ \mu }^{c}\left ( 0 \right )\mid J/\psi \left ( P^{\prime}+q \right )\rangle &=&f_{J/\psi }m_{J/\psi }\epsilon _{\mu },
\end{eqnarray}
 here $f_{J/\psi }$ represents the decay constant of $J/\psi $, $\epsilon _{\mu }$ is the polarization vector of $J/\psi $. The matrix element of the electromagnetic current between $\eta_{c}$ and $J/\psi$ is defined as~\cite{Dudek:2006ej}
\begin{eqnarray}
\label{eq2}
\langle\eta_{c}(P^{\prime})\mid J_{\mu}^{c}(0) \mid J/\psi(P^{\prime}+q)\rangle& = &\epsilon_{\mu abc}P^{\prime a}q^{b}\epsilon^{ c\ast}F_{VP}\frac{2}{m_{J/\psi}+m_{\eta_{c}}},
\end{eqnarray}
where $F_{VP}$ is the form factor of the process $J/\psi(P)\rightarrow\eta_{c}(P^{\prime})+\gamma(q)$.
Substituting the matrix elements Eq.(\ref{eq11}) and (\ref{eq2}) into (\ref{eq1}), the hadronic states representation is converted into the following form
\begin{eqnarray}
\label{eq3}
\Pi _{\mu \nu }\left ( P^{\prime},q \right )&=&2\epsilon _{\mu \nu \alpha \beta }q^{\alpha }P^{\prime\beta }\frac{m_{J/\psi}}{m_{J/\psi}+m_{\eta_{c}}}\frac{f_{J/\psi}F_{VP}}{m_{J/\psi}^{2}-(P^{\prime}+q)^{2}}
+\int_{s_{0}}^{\infty }ds\frac{\rho^{h}(s)}{s-( P^{\prime}+q )^{2}}.
\end{eqnarray}
Note that the first is the contribution from the ground state, and the second is the contributions from the excited states and the continuum.

Having obtained the representation of the correlation function (\ref{eq9}) from the hadronic side, we then calculate it in the deep-Euclidean domain using OPE. In present work, we calculate the leading-order contribution of the correlator Eq.(\ref{eq9}) with the light-cone OPE. Contracting the c-quark fields in Eq.(\ref{eq9}), we obtain free c-quark propagator
\begin{eqnarray}
\label{eq15}
iS\left ( x,0 \right )=i\int\frac{d^{4}k}{\left ( 2\pi  \right )^{4}}e^{-ikx}\frac{\not{k}+m_{c}}{k^{2}-m_{c}^{2}},
\end{eqnarray}
where $m_{c}$ and $k$ are the mass and the four-momentum of c-quark. After replacing corresponding free quark propagator with (\ref{eq15}), transforming $\gamma _{\mu }\gamma _{\alpha }\gamma _{\nu }\rightarrow -i\epsilon _{\mu \alpha \nu \beta }\gamma ^{\beta }\gamma _{5}+...$ in Eq.(\ref{eq9}). we obtain
\begin{eqnarray}
\label{eq14}
\Pi _{\mu \nu }\left ( P^{\prime},q \right )&=&2i\epsilon _{\mu \alpha \nu \beta }\int d^{4}x\int\frac{d^{4}k}{(2\pi)^{4}}e^{i(q-k)x}\frac{k^{\alpha}}{k^{2}-m_{c}^{2}}\langle\eta _{c}\left ( P^{\prime} \right )\mid \bar{C}\left ( x \right )\gamma ^{\beta }\gamma _{5}C\left ( 0 \right )\mid 0\rangle.
\end{eqnarray}

The nonlocal operator in (\ref{eq14}) can be expanded near the light-cone $x^{2}=0$ in components corresponding to different twists. In the leading order of this expansion, at $x^{2}=0$, the matrix element in (\ref{eq14}) has the following parametrization:
\begin{eqnarray}
\langle\eta _{c}\left ( P^{\prime} \right )\mid \bar{C}\left ( x \right )\gamma ^{\beta }\gamma _{5}C\left ( 0 \right )\mid 0\rangle&=-iP^{\prime\beta }f_{\eta _{c}}\int_{0}^{1}due^{iuP^{\prime}x}\phi_{\eta _{c}}\left ( u \right ),
\end{eqnarray}
where $\phi _{\eta _{c}}(u)$ is the light-cone distribution amplitude of $\eta _{c}$ meson, normalized to unity:$\int_{0}^{1}\phi _{\eta _{c}}(u)du=1$. $u$ is longitudinal momentum rate of $c$ quark in $\eta _{c}$ meson, and $f_{\eta _{c}}$ is the decay constant of $\eta _{c}$ meson. Since the distribution amplitude of $\eta_{c}$ meson  is studied comprehensively in~\cite{Lepage:1980fj,Sun:2009zk,Tao:2019rwy}, which is adopted in this work, we do not present their explicit expressions here.
The final expression of the operator product expansion for the correlation function is:
\begin{eqnarray}
\label{eq6}
\Pi _{\mu \nu }\left ( P^{_{\prime}} ,q\right )=2\epsilon _{\mu \nu \alpha \beta }q^{\alpha }P^{\prime\beta }f_{\eta _{c}}\int_{0}^{1}dx\frac{\phi _{\eta _{c}}\left ( x \right )}{m_{c}^{2}-\left ( xP^{\prime}+q \right )^{2}}.
\end{eqnarray}

The function $\rho^{h}(s)$ in the second part of Eq.(\ref{eq3}) is the hadronic spectral density of all excited and continuum states with the same quantum numbers as $J/\psi$ meson. With quark-hadron duality, the hadronic spectral density can be expressed as the following form
\begin{eqnarray}
\label{eqIm}
\int_{s_{0}}^{\infty }ds\frac{\rho^{h}( s )}{s-\left ( P^{\prime}+q \right )^{2}}\simeq \frac{1}{\pi }\int_{s_{0}}^{\infty }ds\frac{{\rm{Im}}\Pi _{\mu \nu }^{(pert)}\left ( P^{\prime},q \right )}{s-\left ( P^{\prime}+q \right )^{2}},
\end{eqnarray}
where
\begin{eqnarray}
\frac{1}{\pi }{\rm{Im}}\Pi_{\mu \nu }^{(pert)}\left ( P^{\prime},q \right )=&2\epsilon _{\mu \nu \alpha \beta }q^{\alpha }P^{\prime\beta }f_{\eta _{c}}\int_{0}^{1}dx\phi _{\eta _{c}}\left ( x \right )\delta \left ( m_{c}^{2}\right.
\left.+x\bar{x}P^{\prime2}-\bar{x}q^{2}-xs_{0} \right ),
\end{eqnarray}
with $\bar{x}=1-x$. Then Eq.(\ref{eqIm}) is transformed into
\begin{eqnarray}
\frac{1}{\pi }\int_{s_{0}}^{\infty }ds\frac{{\rm{Im}}\Pi _{\mu \nu }^{(pert)}\left ( P^{\prime},q \right )}{s-\left ( P^{\prime}+q \right )^{2}}\simeq 2\epsilon _{\mu \nu \alpha \beta }q^{\alpha }P^{\prime\beta }f_{\eta _{c}}\int_{0}^{\Delta}dx\frac{\phi _{\eta _{c}}\left ( x \right )}{m_{c}^{2}-\left ( xP^{\prime}+q \right )^{2}},
\end{eqnarray}
where $\Delta $ is the solution of equation $m_{c}^{2}+x\bar{x}P^{\prime2}-\bar{x}q^{2}-xs_{0}=0$ in the range $[0,1]$, whose explicit form is shown in Eq.(\ref{eqDelta}).

Finally, the hadronic states representation in Eq.(\ref{eq3}) is matched with OPE in Eq.(\ref{eq6}). In order to suppress the excited states and the continuum and to boost the contribution of the ground state $J/\psi$, one perform the Borel transformations over the variables $(P'+q)^{2}$ and $(xP'+q)^{2}$ on both the hadronic side and the OPE side~\cite{Shifman:1978bx,Shifman:1978by}
\begin{eqnarray}
B_{M^{2}}\frac{1}{m_{J/\psi }^{2}-\left ( P^{\prime}+q \right )^{2}}&=&\frac{1}{M^{2}}e^{-\frac{m_{J/\psi }^{2}}{M^{2}}},\nonumber\\
B_{M^{2}}\frac{1}{m_{c }^{2}-\left ( xP^{\prime}+q \right )^{2}}&=&\frac{1}{xM^{2}}e^{\left \{ -\frac{1}{xM^{2}}\left [ m_{c }^{2}+x\left ( 1-x \right )P^{\prime2}-\left ( 1-x \right )q^{2} \right ] \right \}},
\end{eqnarray}
where $M^{2}$ is the Borel parameter. Then we obtain the LCSR for the form factor $F_{VP}$ at $q^{2}=-Q^{2}=0$
\begin{eqnarray}
\label{eq5}
F_{VP}=\frac{f_{\eta _{c}}}{f_{J/\psi}}\frac{m_{J/\psi}+m_{\eta_{c}}}{m_{J/\psi}}\int_{\Delta}^{1}dx\frac{\phi _{\eta _{c}\left ( x \right )}}{x}e^{\left \{ -\frac{1}{xM^{2}}\left [ m_{c }^{2}+x\left ( 1-x \right )m_{\eta_{c}}^{2}-\left ( 1-x \right )q^{2} \right ] +\frac{m_{J/\psi}^{2}}{M^{2}}\right \}},
\end{eqnarray}
in Eq.(\ref{eq5}), the lower limit in integral is

\begin{eqnarray}
\label{eqDelta}
\Delta=\frac{1}{2m_{\eta _{c}}^{2}}\left [ \sqrt{\left ( s_{0}-m_{\eta _{c}}^{2}+Q^{2} \right )^{2}+4\left ( m_{c}^{2}+ Q^{2}\right )m_{\eta _{c}}^{2}}-\left ( s_{0}-m_{\eta _{c}}^{2}-q^{2} \right ) \right ],
\end{eqnarray}
where the four-momentum of the $\eta _{c}$ meson satisfies $P'^{2}=m_{\eta_{c}}^{2}$.

B.Decay width

Generally, one can easily get the expression of the decay width of $J/\psi(P)\rightarrow\eta_{c}(P^{\prime})+\gamma(q)$
\begin{eqnarray}
\label{eq8}
\Gamma =\int d\Omega_{q}\frac{1}{32\pi^{2}}\frac{|\overrightarrow{q}|}{m_{J/\psi}^{2}}|\overline{\mathcal{M}}|^{2},
\end{eqnarray}
where
 \begin{eqnarray}
 \label{eq12}
  |\overrightarrow{q}|=\frac{m_{J/\psi}^{2}-m_{\eta_{c}}^{2}}{2m_{J/\psi}}
\end{eqnarray}
  presents the three-momentum of the final photon, which is determined by the energy-momentum conservation in the rest frame of the initial $J/\psi$ meson, $|\overline{\mathcal{M}}|^{2}$ is related to scattering matrix element $\mathcal{M}$ by\\
\begin{eqnarray}
\label{eq13}
|\overline{\mathcal{M}}|^{2}=\frac{1}{2J+1}\sum _{m_{j},\lambda}|\mathcal{M}_{m_{j},\lambda}|^{2},
\end{eqnarray}
where $J$ is the spin of the initial particle $J/\psi$, $m_{j}$ and $\lambda$ stand for the polarizations of $J/\psi$ meson and the photon respectively. Eq.(\ref{eq13}) contains sum over the final photon polarization and average over the initial $J/\psi$ polarization.

 The transition amplitude of process $J/\psi\rightarrow\eta_{c}+\gamma$ is defined in terms of the standard Lorentz covariant decomposition
\begin{eqnarray}
\mathcal{M}_{m_{j},\lambda}=\langle\eta_{c}(P^{\prime})\mid J^{\mu}_{em}(0)\mid J/\psi(P,m_{j})\rangle\epsilon_{\mu,\lambda}^{\ast}|_{Q^{2}=0},
\end{eqnarray}
with the Minkowski space-time matrix element for this transition~\cite{Dudek:2006ej}
\begin{eqnarray}
\langle \eta_{c}\left ( P^{\prime} \right )\mid J_{em }^{\mu}\left ( 0 \right )\mid J/\psi \left ( P^{\prime}+q \right )\rangle =\frac{2W_{VP}}{m_{J/\psi}+m_{\eta_{c}}}\epsilon^{\mu} _{abc}P^{\prime a}q^{b}\epsilon^{c},
\end{eqnarray}
here $W_{VP}$ is real and time-reversal invariant.

If the sea quark contributions can be neglected, one can use $J_{em }^{\mu}(0)=Q_{c}e\overline{C}\gamma_{\mu}C$ as an approximate electromagnetic vector current for the electromagnetic decay $J/\psi\rightarrow\eta_{c}+\gamma$. The dimensionless fractional charge for the charm quark is $ Q_{c} = +2/3$, and the electric charge is $e =\sqrt{4\pi\alpha_{em}}$. Based on the OZI rule, we neglect contributions from the quark annihilation diagrams and only consider the contributions of connected diagrams. For charmonium, based on the charge conjugation symmetry, the antiquark gives the same contribution as the quark for the total hadronic current. So, we calculate the hadron matrix element from the charmonium quark part. Thus we compute $W_{VP}$ which is related to $F_{VP}$ by
\begin{eqnarray}
W_{VP}=2\times\frac{2}{3}e\times F_{VP} ,
\end{eqnarray}
  the factor 2 comes from the insertion of the electromagnetic current to both the quark and antiquark lines. Substituting $\left |\overline{\mathcal{M}}\right |^{2}$ into Eq.(\ref{eq8}), integrating the phase space, after some tedious calculations, we get~\cite{Dudek:2006ej}
\begin{eqnarray}
\Gamma=\alpha_{em}\frac{(m_{J/\psi}^{2}-m_{\eta_{c}}^{2})^{3}}{(2m_{J/\psi})^{3}(m_{J/\psi}+m_{\eta_{c}})^{2}}\frac{64}
{27}|F_{VP}|^{2},
\end{eqnarray}
where $F_{VP}$ is the radiative transition form factor at $Q^{2}=0$ as in Eq.(\ref{eq5}).

\section{Numerical analysis}\label{sec3}

In this section we focus on the numerical study of the form factor and decay width for $J/\psi\rightarrow\eta_{c}+\gamma$. For this purpose, we first take several parameters as $m_{\eta _{c}}=2.9835\ \mathrm{GeV}$, $m_{J/\psi}=3.0969\ \mathrm{GeV}$ and $f_{\eta_{c}}/f_{J/\psi}=0.54 $~\cite{Anashin:2014wva,Edwards:2000bb,Deshpande:1994mk}.

In addition to the above input parameters, the form factor $F_{VP}$ also contains three extra parameters. These parameters are the Borel parameter $M^{2}$, the threshold $s_{0}$, and the mass of the $c$ quark. According to the light-cone sum rules approach, we determine the working regions of the parameters $M^{2}$ and $s_{0}$. In order to suppress the contributions of the excited states and the continuum, and the contributions of the higher twist distribution amplitudes of $\eta_{c}$ meson, the Borel parameter $M^{2}$ should be in the region $15\ \mathrm{GeV^{2}}\leq M^{2}\leq 25\ \mathrm{GeV^{2}}$. It can be seen from Fig.1 that the error in the form factor caused by Borel parameter is very small, about 2.6\%. For convenience, we take Borel parameter as $20\ \mathrm{GeV^{2}}$.

The threshold is not totally arbitrary, which is chosen at the value which separates the ground state from the excited states and the continuum. We take its value within the region $s_{0}=(3.60^{2}-3.70^{2})\ \mathrm{GeV^{2}}$. In order to judge the dependence of the form factor on threshold parameter $s_{0}$, we present the form factors at several discrete values of the continuum threshold $s_{0}=3.60^{2}\ \mathrm{GeV^{2}}, s_{0}=3.65^{2}\ \mathrm{GeV^{2}}$ and $s_{0}=3.70^{2}\ \mathrm{GeV^{2}}$ in Fig.1. As is seen from Fig.1, the dependence of the form factor on threshold parameter $s_{0}$ is relatively small. In our calculation, we adopt the threshold parameter $s_{0}=3.65^{2}\ \mathrm{GeV^{2}}$.

\begin{figure}[!htbp]
\centering
\begin{tabular}{c}
\includegraphics[width=0.5\textwidth]{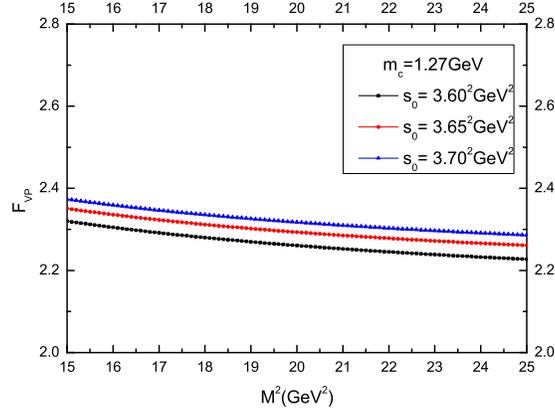}
\end{tabular}
\caption{(color online) The dependence of the form factor for radiative transition $J/\psi\rightarrow\eta_{c}+\gamma$ on the Borel parameter $M^{2}$ at $m_{c}=1.27\ \mathrm{GeV}$ and discrete values of the threshold $s_{0}$.  }\label{potentialshape}
\end{figure}

\begin{figure}[!htbp]
\centering
\begin{tabular}{c}
\includegraphics[width=0.5\textwidth]{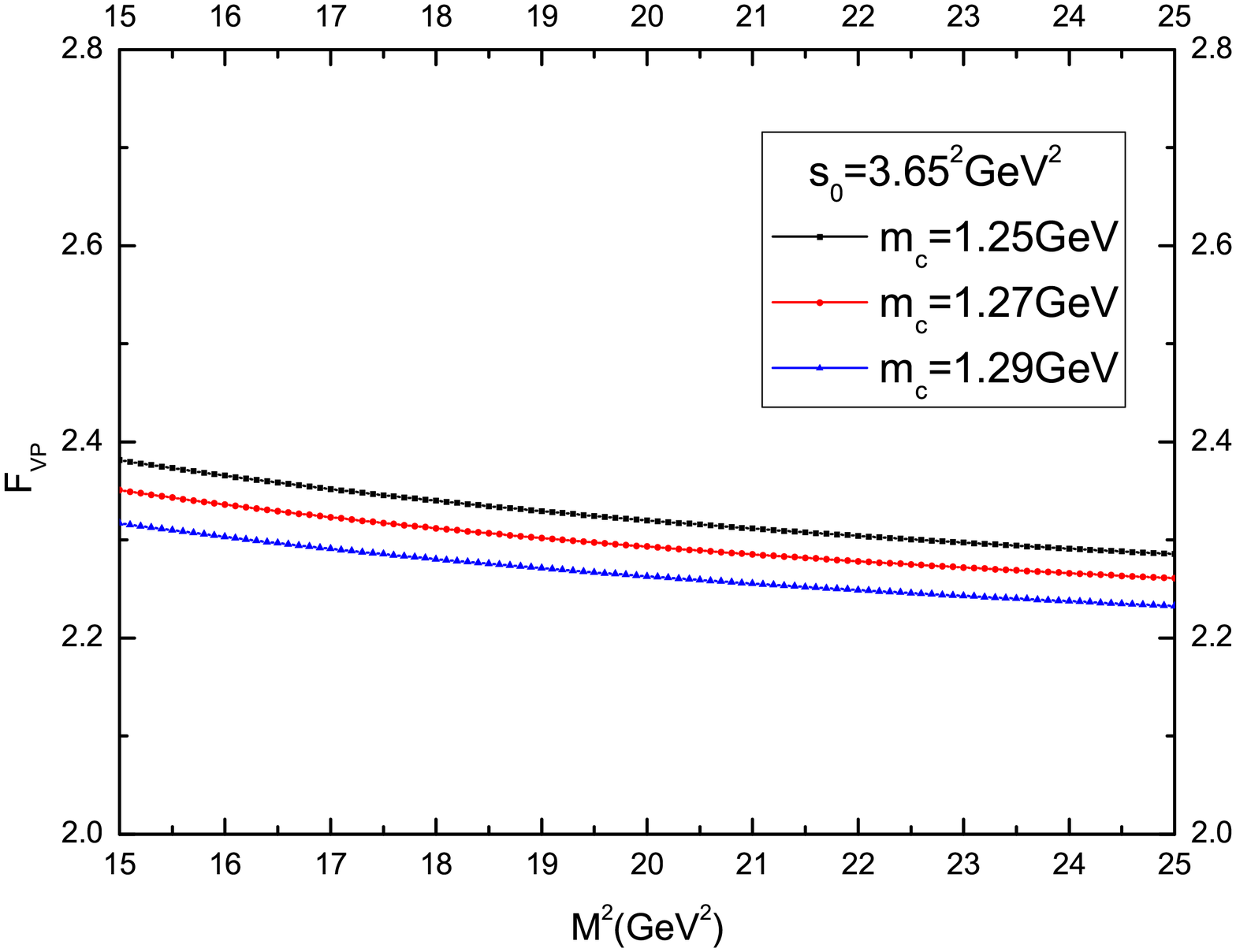}
\end{tabular}
\caption{(color online) The dependence of the form factor for radiative transition $J/\psi\rightarrow\eta_{c}+\gamma$ on the Borel parameter $M^{2}$ at the threshold $s_{0}=3.65^{2}\ \mathrm{GeV}^{2}$ and discrete values of the c quark mass.  }\label{potentialshape}
\end{figure}
\begin{figure}[!htbp]
\centering
\begin{tabular}{c}
\includegraphics[width=0.5\textwidth]{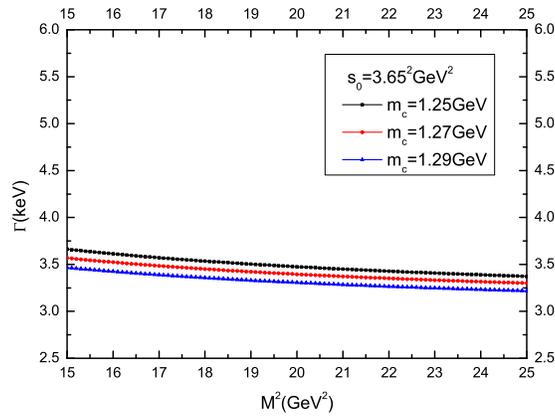}
\end{tabular}
\caption{(color online) The dependence of the decay width for radiative transition $J/\psi\rightarrow\eta_{c}+\gamma$ on the Borel parameter $M^{2}$ at the threshold $s_{0}=3.65^{2}\ \mathrm{GeV}^{2}$ and discrete values of the c quark mass.}\label{potentialshape}
\end{figure}
Having fixed the working regions of $M^{2}$ and $s_{0}$, we now determine the mass of $c$ quark. In PDG of 2018~\cite{Tanabashi:2018oca}, the mass of $c$ quark is taken as $m_{c}=1.27\pm0.02\ \mathrm{GeV}$. Fig.2 and Fig.3 show the form factor and the decay width at three discrete masses $m_{c}=1.25\ \mathrm{GeV}, m_{c}=1.27\ \mathrm{GeV}$ and $m_{c}=1.29\ \mathrm{GeV}$ respectively. As can be seen from the figures, the values of the form factor and the decay width decrease with the increase of the mass of $c$ quark. The error caused by the $c$
quark mass is one of the main uncertainties in our results. In order to compare with the experimental data, we take the central value of the mass of $c$ quark as $m_{c}=1.27\ \mathrm{GeV}$.

\renewcommand\tabcolsep{0.53cm}
\renewcommand{\arraystretch}{1.5}
\begin{table*}[!htbp]
\caption{Comparison of decay width of our method with experiments and other methods.}\label{Table I}
\begin{tabular}{|c|c|c|}
\hline\hline
 $\Gamma\ \mathrm{(keV)}$ &  Ref. & comments
\\ \hline
$1.14$ & $\mathrm{JG}85$~\cite{Gaiser:1985ix} & Crystal Ball
\\
$1.84$ & $\mathrm{RE}08$~\cite{Mitchell:2008aa} & CLEO Collaboration
\\
$2.98\pm 0.18_{-0.33}^{+0.15}$ & $\mathrm{A14}$~\cite{Anashin:2014wva} & KEDR
\\ \hline
$2.9$ & $\mathrm{DER}06$~\cite{Dudek:2006ej} & (quenched) Lattice QCD
\\
$2.51(8)$ & $\mathrm{DET}09$~\cite{Dudek:2009kk}& (quenched) Lattice QCD
\\
$2.47$ & $\mathrm{GDCY}19$~\cite{Gui:2019dtm} &  (quenched) Lattice QCD
\\
$2.49\pm0.19 $ & $\mathrm{DDDF}$12~\cite{Donald:2012ga} & ($N_{f}=2+1$) Lattice QCD
\\
$2.84(6)$ & $\mathrm{C}$11~\cite{Chen:2011kpa} & ($N_{f}=2$) Lattice QCD
\\
$2.64(11)(3)$ & $\mathrm{BS}$13~\cite{Becirevic:2012dc} & ($N_{f}=2$) Lattice QCD
\\ \hline
$1.5\pm 1.0$ & $\mathrm{BGV}$06~\cite{Brambilla:2005zw} & non-relativistic effective field theories of QCD
\\
$2.12(40)$ & $\mathrm{PS}$13~\cite{Pineda:2013lta} & potential non-relativistic effective field theories of QCD
\\ \hline
$1.05$ & $\mathrm{EFG}$03~\cite{Ebert:2002pp} & relativistic quark model
\\
$2.39$ & $\mathrm{DXGZ}$15~\cite{Deng:2016stx} & constituent quark model
\\
$2.9$ & $\mathrm{BGS}$05~\cite{Barnes:2005pb} & non-relativistic potential model
\\
$2.9$ & $\mathrm{GYS}$14~\cite{Guo:2014zva} & Coulomb gauge approach
\\ \hline
$1.7-3.3(0.75-1.0)$ & $\mathrm{EFG}$03~\cite{Khodjamirian:1979fa} & dispersion sum rules
\\
$2.1\pm0.4$ & $\mathrm{TM}$84~\cite{Aliev:1985vt} & QCD sum rules
\\
$2.6\pm0.5$ & $\mathrm{BR}$85~\cite{Beilin:1984pf} & QCD sum rules
\\
$3.39\pm0.34$ & $\mathrm{This \ work}$ & light cone sum rules
\\ \hline \hline
\end{tabular}
\end{table*}
\medskip
Table I presents a comparison of our decay width with those from experiments and other theoretical models for this process. The first column shows the experimental values of decay width from Crystal Ball~\cite{Gaiser:1985ix}, CLEO Collaboration~\cite{Mitchell:2008aa} and KEDR~\cite{Anashin:2014wva}. It can be seen that the experimental results in different periods of time are inconsistent, and the values are gradually increasing. We hope more efforts on the experimental field will be made to clarify the inconsistencies among various experiments. The second column is the results of lattice QCD~\cite{Dudek:2006ej,Dudek:2009kk,Gui:2019dtm,Donald:2012ga,Chen:2011kpa,Becirevic:2012dc}, which are totally independent of the mass of $c$ quark, and the lattice QCD results $\mathrm{2.4-2.9\ keV}$ keep good consistent with current experimental value~\cite{Anashin:2014wva}. The third column is the outcomes from the non-relativistic effective field theories of QCD~\cite{Brambilla:2005zw,Pineda:2013lta}, with Ref.~\cite{Brambilla:2005zw} in the strict weak-coupling limit of potential NRQCD. In the fourth column, the result of relativistic quark model~\cite{Ebert:2002pp} mixtureing scalar potential and vector potential is consistent with the experimental value~\cite{Gaiser:1985ix} at that time.
 The fifth column is the results of dispersion sum rules~\cite{Khodjamirian:1979fa}, QCD sum rules~\cite{Beilin:1984pf,Aliev:1985vt}and LCSR. Here, the result of the dispersion sum rules~\cite{Khodjamirian:1979fa} was obtained on the basis of local duality hypothesis, when the mass of $c$ quark is taken as $m_{c}=1.25(1.30)\ \mathrm{GeV}$, the decay width is $1.7-3.3\ \mathrm{keV}(0.75-1.0 \ \mathrm{keV})$. The result from the QCD sum rules~\cite{Beilin:1984pf} is $2.6\pm0.5\ \mathrm{keV}$ when $m_{c}=1.26\ \mathrm{GeV}$, the author further pointed out that the decay width value corresponding to $m_{c}=1.26\ \mathrm{GeV}$ is $40\%$ higher than that for $m_{c}=1.28\ \mathrm{GeV}$. The last in the fifth column is our result with LCSR, $\Gamma=3.39\pm0.34\ \mathrm{keV}$. The error $(10-15)\%$ in our result is caused by the uncertainties of the input parameters $M^{2},\ s_{0},\ m_{c}$ and quark hadron duality ansatz. Among them, the primary source of errors comes from the uncertainty of $m_{c}$ and quark hadron duality ansatz. It can be seen that the outcomes from kinds of the sum rules depend on the mass of $c $ quark.

  The results in the fifth column are roughly coincident and all based on the QCD sum rules method, which has a very close relationship with QCD that is lack in the potential model~\cite{Ebert:2002pp,Deng:2016stx,Barnes:2005pb,Guo:2014zva} in the fourth column. For the M1 radiative transitions, the decay widths are sensitive to relativistic effects. The LCSR using relativistic wave function can reflect the relativistic nature of $J/\psi\rightarrow\eta_{c}+\gamma$ process, which is deficient in non-relativistic effective field theories of QCD~\cite{Brambilla:2005zw,Pineda:2013lta} in the third column and the QCD sum rules~\cite{Beilin:1984pf} as well as the dispersion sum rules~\cite{Khodjamirian:1979fa} in the fifth column. The point is consistent with that mentioned in potential NRQCD~\cite{Brambilla:2005zw} in which the author clearly indicates large relativistic
  corrections.

\section{summary}\label{sec5}

We have studied the radiative transition of charmonium $J/\psi\rightarrow\eta_{c}+\gamma$ within the LCSR. For this purpose, the transition form factor of the relevant decay is calculated.
Using the value of the relevant form factor at $Q^{2}=0$, the corresponding decay width is estimated.
Here, the form factor is $F_{VP}=2.293$ and decay width is $\Gamma=3.39\pm0.34\ \mathrm{keV}$. The error of our results mainly comes from the uncertainty of $c$ quark mass and quark hadron duality ansatz. Our result is compatible with recent experimental date~\cite{Anashin:2014wva} within errors, and it is consistent with the results of previous sum rules method~\cite{Khodjamirian:1979fa,Beilin:1984pf}.

It should be pointed out that the relativistic effect for the process $J/\psi\rightarrow\eta_{c}+\gamma$ is significant~\cite{Ebert:2002pp,Brambilla:2005zw}, thus it is reasonable to apply the LCSR method to the treatment of this process, since the light-cone distribution amplitude could reflect the relativistic effect.                      However, there are limitations in our approach. In this paper, only the contribution of the leading twist distribution amplitude of $\eta_{c}$ meson is considered. If the contributions of the higher twist distribution amplitudes are also taken into account, our result will be more reliable. The LCSR approach can also be used to study other radiative transitions between charmoniums such as $\psi'\rightarrow\eta'_{c}\gamma,\psi'\rightarrow\eta_{c}\gamma,\eta'_{c}\rightarrow J/\psi\gamma$ and etc., the corresponding work will be carried out in the future.

\section*{ACKNOWLEDGEMENTS}

This work was supported in part by National Natural Science Foundation of China under Grant No.11365018 and No.11375240.


\begin{thebibliography}{99}
\bibitem{Gaiser:1985ix}
  J.~Gaiser {\it et al.},
  Phys.\ Rev.\ D {\bf 34}, 711 (1986).
\bibitem{Mitchell:2008aa}
  R.~E.~Mitchell {\it et al.} [CLEO Collaboration],
  Phys.\ Rev.\ Lett.\  {\bf 102}, 011801 (2009)
  Erratum: [Phys.\ Rev.\ Lett.\  {\bf 106}, 159903 (2011)]
  [arXiv:0805.0252].
\bibitem{Anashin:2014wva}
  V.~V.~Anashin {\it et al.},
  Phys.\ Lett.\ B {\bf 738}, 391 (2014)
  [arXiv:1406.7644].
\bibitem{Khodjamirian:1979fa}
  A.~Y.~Khodjamirian,
  Phys.\ Lett.\  {\bf 90B}, 460 (1980).
\bibitem{Beilin:1984pf}
  V.~A.~Beilin and A.~V.~Radyushkin,
  Nucl.\ Phys.\ B {\bf 260}, 61 (1985).
\bibitem{Ebert:2002pp}
  D.~Ebert, R.~N.~Faustov and V.~O.~Galkin,
  Phys.\ Rev.\ D {\bf 67}, 014027 (2003)
  [hep-ph/0210381].
\bibitem{Deng:2016stx}
  W.~J.~Deng, H.~Liu, L.~C.~Gui and X.~H.~Zhong,
  Phys.\ Rev.\ D {\bf 95}, no. 3, 034026 (2017)
  [arXiv:1608.00287].
\bibitem{Barnes:2005pb}
  T.~Barnes, S.~Godfrey and E.~S.~Swanson,
  Phys.\ Rev.\ D {\bf 72}, 054026 (2005)
  [hep-ph/0505002].
\bibitem{Guo:2014zva}
  P.~Guo, T.~Y¨¦pez-Mart¨ªnez and A.~P.~Szczepaniak,
  Phys.\ Rev.\ D {\bf 89}, no. 11, 116005 (2014)
  [arXiv:1402.5863].
\bibitem{Brambilla:2005zw}
  N.~Brambilla, Y.~Jia and A.~Vairo,
  Phys.\ Rev.\ D {\bf 73}, 054005 (2006)
  [hep-ph/0512369].
\bibitem{Pineda:2013lta}
  A.~Pineda and J.~Segovia,
  Phys.\ Rev.\ D {\bf 87}, no. 7, 074024 (2013)
  [arXiv:1302.3528].
\bibitem{Dudek:2006ej}
  J.~J.~Dudek, R.~G.~Edwards and D.~G.~Richards,
  Phys.\ Rev.\ D {\bf 73}, 074507 (2006)
  [hep-ph/0601137].
\bibitem{Dudek:2009kk}
  J.~J.~Dudek, R.~Edwards and C.~E.~Thomas,
  Phys.\ Rev.\ D {\bf 79}, 094504 (2009)
  [arXiv:0902.2241].
\bibitem{Gui:2019dtm}
  L.~C.~Gui, J.~M.~Dong, Y.~Chen and Y.~B.~Yang,
  Phys.\ Rev.\ D {\bf 100}, no. 5, 054511 (2019)
  [arXiv:1906.03666].
\bibitem{Donald:2012ga}
  G.~C.~Donald, C.~T.~H.~Davies, R.~J.~Dowdall, E.~Follana, K.~Hornbostel, J.~Koponen, G.~P.~Lepage and C.~McNeile,
  Phys.\ Rev.\ D {\bf 86}, 094501 (2012)
  [arXiv:1208.2855].
\bibitem{Chen:2011kpa}
  Y.~Chen {\it et al.},
  Phys.\ Rev.\ D {\bf 84}, 034503 (2011)
  [arXiv:1104.2655].
\bibitem{Becirevic:2012dc}
  D.~Becirevic and F.~Sanfilippo,
  JHEP {\bf 1301}, 028 (2013)
  [arXiv:1206.1445].
\bibitem{Shifman:1978bx}
  M.~A.~Shifman, A.~I.~Vainshtein and V.~I.~Zakharov,
  Nucl.\ Phys.\ B {\bf 147}, 385 (1979).
\bibitem{Colangelo:2000dp}
  P.~Colangelo and A.~Khodjamirian,
  In *Shifman, M. (ed.): At the frontier of particle physics, vol. 3* 1495-1576
  [hep-ph/0010175].
\bibitem{Sun:2009zk}
  Y.~J.~Sun, X.~G.~Wu, F.~Zuo and T.~Huang,
  Eur.\ Phys.\ J.\ C {\bf 67}, 117 (2010)
  [arXiv:0911.0963].
\bibitem{Tao:2019rwy}
  H.~J.~Tao, Y.~J.~Sun, S.~P.~Guo, W.~Hong and Q.~Huang,
  arXiv:1901.09142.
\bibitem{Lepage:1980fj}
  G.~P.~Lepage and S.~J.~Brodsky,
  Phys.\ Rev.\ D {\bf 22}, 2157 (1980).
\bibitem{Shifman:1978by}
  M.~A.~Shifman, A.~I.~Vainshtein and V.~I.~Zakharov,
  Nucl.\ Phys.\ B {\bf 147}, 448 (1979).
\bibitem{Edwards:2000bb}
  K.~W.~Edwards {\it et al.} [CLEO Collaboration],
  Phys.\ Rev.\ Lett.\  {\bf 86}, 30 (2001)
  [hep-ex/0007012].
\bibitem{Deshpande:1994mk}
  N.~G.~Deshpande and J.~Trampetic,
  Phys.\ Lett.\ B {\bf 339}, 270 (1994)
  [hep-ph/9406393].
\bibitem{Tanabashi:2018oca}
  M.~Tanabashi {\it et al.} [Particle Data Group],
  Phys.\ Rev.\ D {\bf 98}, no. 3, 030001 (2018).
\bibitem{Aliev:1985vt}
  T.~M.~Aliev,
  Z.\ Phys.\ C {\bf 26}, 275 (1984).
\end{thebibliography}
\end{document}